\documentclass[fleqn,usenatbib]{mnras}

\usepackage{newtxtext,newtxmath}

\usepackage[T1]{fontenc}

\DeclareRobustCommand{\VAN}[3]{#2}
\let\VANthebibliography\thebibliography
\def\thebibliography{\DeclareRobustCommand{\VAN}[3]{##3}\VANthebibliography}

\usepackage{graphicx}	
\usepackage{amsmath}	

\usepackage{xspace}
\usepackage{xcolor}

\title[CO Bands in CP Cep]{Changing CO Bands in Near-IR Spectra of CP Cephei}

\author[S. Call et al.]{
Scott G. Call,$^{1}$
Eric G. Hintz,$^{1}$
Steve Ardern,$^{2}$
Victoria Scowcroft,$^{2}$
and Timothy D. Morrell$^{1}$
\\
$^{1}$Department of Physics and Astronomy, Brigham Young University, Provo, Utah, 84602, USA\\
$^{2}$Department of Physics, University of Bath, Claverton Down, Bath, BA2 7AY, UK\\
}

\date{Accepted XXX. Received YYY; in original form ZZZ}

\pubyear{2024}

\begin{document}
\label{firstpage}
\pagerange{\pageref{firstpage}--\pageref{lastpage}}
\maketitle

\begin{abstract}
We present time-series near-infrared spectra for the classical Cepheid, CP~Cephei, from the Astrophysical Research Consortium 3.5-m telescope and near-infrared spectrograph, \emph{TripleSpec}, at Apache Point Observatory, NM, USA. Spectral observations were made at nine points through the minimum and partway up the ascending portion of the optical light curve for the star. Carbon monoxide (CO) was detected in absorption in the 2.3-$\mu$m region for each observation. We measured the strength of CO absorption using the 2-0 band head in the feature for each observation and confirm that the CO varies with pulsation. We show that these measurements follow the $(J-K)$ colour curve, confirming that temperature drives the destruction of CO. By obtaining convolved filter magnitudes from the spectral data we found that the effect of the CO feature on $K$ magnitudes is small, unlike the CO feature in the mid-infrared at 4.5-$\mu$m. The dissociation of CO in the near-infrared spectra tracks with the effect seen in the mid-infrared photometric measurements of a similar Galactic Cepheid. Confirmation of the varying CO feature illustrates the need for further investigations into the related mid-infrared period-colour-metallicity relation in order to address the impact of Cepheid metallicities on the Hubble tension.
\end{abstract}




\section{Introduction}
\label{sec:intro}

The Leavitt law (previously known as the period-luminosity relation) for classical Cepheid variables has been shown to have decreased dispersion at longer wavelengths \citep{madore91}. As we move beyond the optical region, temperature variation of the stars and extinction due to the interstellar medium become smaller obstacles to obtaining accurate measurements. The near and mid-infrared (NIR and MIR) have become the priority regions for distance measurements using the Leavitt law \citep[see e.g.][and references therein]{Freedman2012, riess22}. 

The conditions present in supergiant stars allow carbon monoxide (CO) to form in their atmospheres \citep[see][]{lancon92}. \cite{marengo10} found that variable CO was the cause of colour variations in the $4.5$-$\mathrm{\mu m}$ passband for a selection of Cepheids, and modeled the CO variation at different phases for one star. The dissociation and recombination of CO in Cepheid atmospheres is the result of changes in reaction kinetics with temperature and thus, correlated with pulsation phase \citep[discussions of this are given in][]{scowcroft11,monson12,scowcroft16}. As temperature decreases and CO forms, the flux in the associated passband (e.g. $K$-band in the NIR, or \textit{JWST} F444W in the MIR) will be suppressed due to greater absorption \citep[see Figure 2 of][]{scowcroft16}.

In addition to the photometric detections in the MIR, \cite{hamer} observed CO in four classical Cepheids at sub-mm wavelengths. This represents the first direct evidence of CO in Cepheid atmospheres. Here we present NIR spectroscopic data for CP~Cephei (hereafter CP Cep), one of the targets observed by \cite{hamer}, confirming variation of CO via the 2.3-$\mathrm{\mu m}$ band heads. Our observations cover over 40\% of the period of the star, and the strength of CO in these observations confirm the variation is correlated with pulsation of the star.

This work contains spectroscopic observations of temperature-dependent CO in CP~Cep. In Section~\ref{sec:observations} we discuss the instrumentation and data reduction and provide context for the observations. Our analyses of changes in CO strength and the results from the NIR observations are given in Section~\ref{sec:results}, with Section~\ref{sec:discussion} comparing our observations to predictions from the MIR colour curve of a similar Galactic Cepheid. We conclude with a summary of our findings in Section ~\ref{sec:conclusions}.

\section{Observations}
\label{sec:observations}

Nine NIR observations of CP~Cep were obtained using the Astrophysical Research Consortium 3.5-m telescope at Apache Point Observatory between the beginning of July and the middle of October, 2023. The NIR spectrograph, \emph{TripleSpec}, covers a range of $0.95 {\text -} 2.46\, \mathrm{\mu m}$ with resolving power $\mathrm{R} = 3500$ \citep{wilson04}. Observations were ``nodded'', or dithered, between two points on the slit allowing for subtraction of atmospheric emission lines and background. Standard stars of spectral type A0 were observed close in time and airmass ($X = \mathrm{sec}(z)$ where $z$ is the zenith angle) to CP~Cep for flux calibration and correction of telluric absorption. 

\begin{figure*}
    \includegraphics[width=0.7\textwidth]{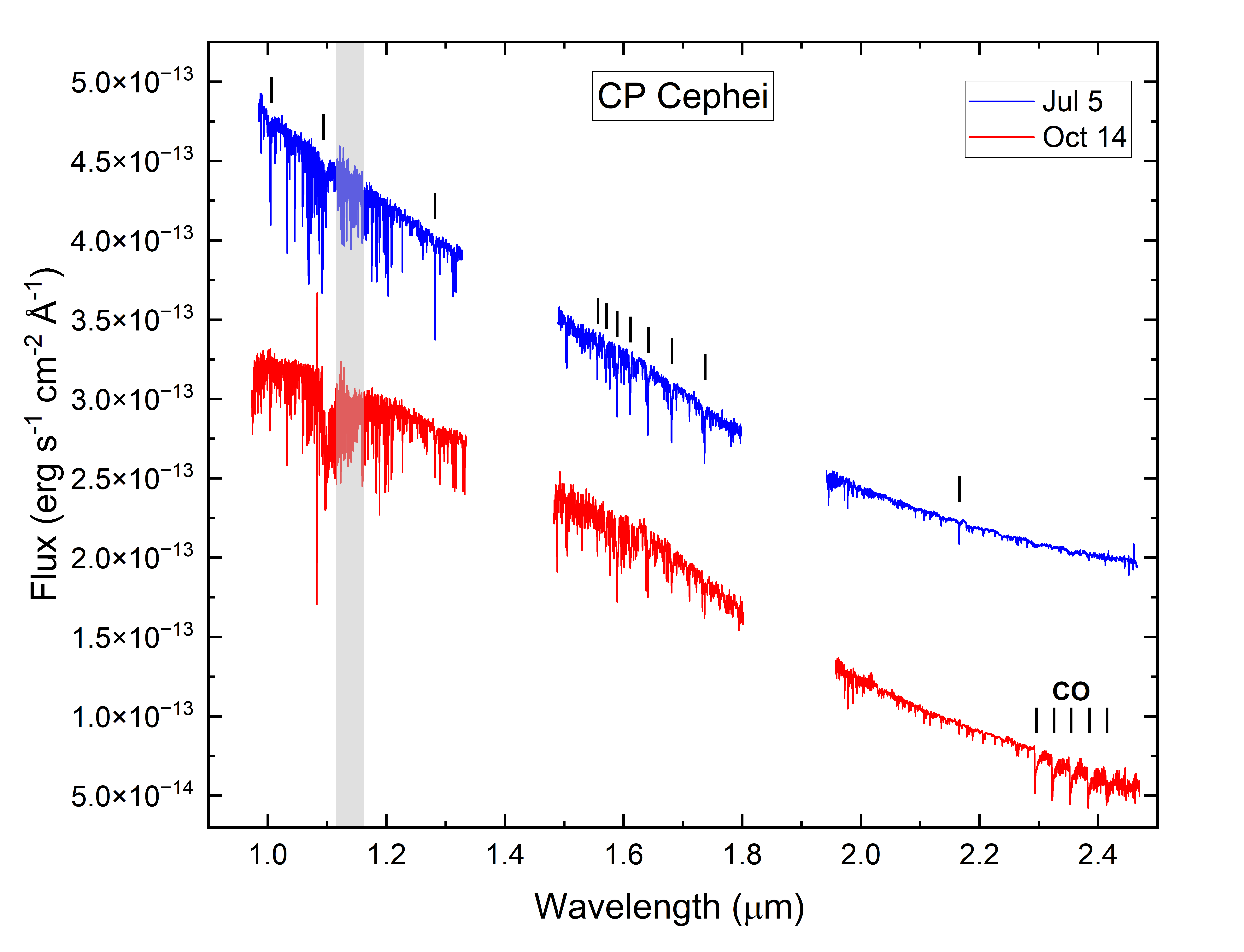}
    \caption{Calibrated spectra for CP~Cep for two nights of observation; July 5th (CO minimum) and October 14th (CO maximum). Low signal-to-noise segments due to atmospheric absorption were removed, with the exception of the gray shaded region near 1.1 $\mu\mathrm{m}$. Major hydrogen lines are shown by black lines above the July 5th spectra, and CO bands are labeled in the October 14th spectra. The July 5th spectra is offset by $+1.5 \times 10^{-13}\, \mathrm{erg\, s^{-1}\, cm^{-2}}$ \r{A}$^{-1}$ for visualization purposes.}
    \label{fig:fluxes}
\end{figure*}

\begin{figure}
    \includegraphics[width=\columnwidth]{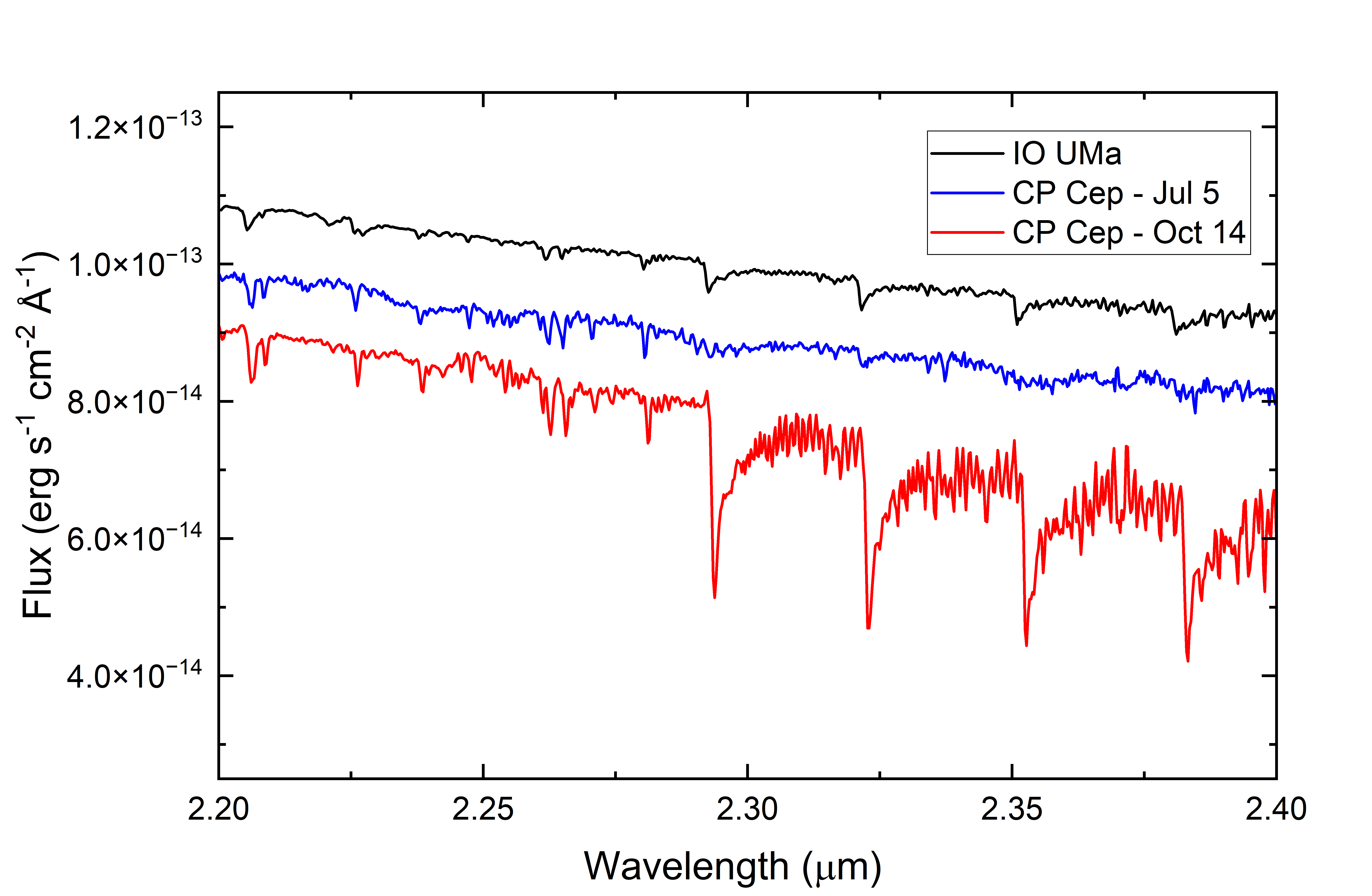}
    \caption{A zoom into the CO region for the two observations in Figure \ref{fig:fluxes}. Spectra from eclipsing binary system, IO Ursae Majoris, is included, which has CO features due to the cool secondary component \citep{soydugan13}. The July 5th observation was shifted $+3.0 \times 10^{-14}\, \mathrm{erg\, s^{-1}\, cm^{-2}}$ \r{A}$^{-1}$, and IO UMa $+5.0 \times 10^{-14}\, \mathrm{erg\, s^{-1}\, cm^{-2}}$ \r{A}$^{-1}$ for visualization purposes.}
    \label{fig:CO-comp}
\end{figure}

Table \ref{tab:obs} details the observations and sky conditions. The spectral extraction and telluric correction were performed using a modified version of \emph{SpexTool} \citep{cushing04, vacca03}. The telluric correction is expectedly better at higher atmospheric transmission regions (i.e. middle of $J, H, \mathrm{and}\,K$ passbands). For our data, this is not only because less correction is necessary, but the wavelength calibration near the strongest telluric absorption is not ideal. When comparing our data to the transmission profile on the edges of the orders, the wavelengths of atmospheric absorption features would be offset slightly which lead to over-correction and increased noise. This can be seen in the region beyond 2.4-$\mathrm{\mu m}$ in Figure \ref{fig:fluxes}.

We had a small window to observe CP~Cep at an acceptable airmass ($X < 1.5$) on July 4th. The target was higher in the sky during our observing run the following night, with almost 26 hours between observations. The sky conditions on both July nights were mostly clear with the possibility of thin clouds affecting the strength of signal. For the August observation, clouds were intermittent at the beginning of the session which led to a larger difference in airmass between our standard and CP~Cep. The October observations were taken over six consecutive nights at roughly the same time each night. Thin clouds may have been present during the October 18th observing session, but the other five nights were clear. Flux values should not be considered absolute due to uncertainties stemming from sky conditions.

Two examples of NIR spectra for CP~Cep are given in Figure \ref{fig:fluxes}, with the July 5th spectra shifted to enable comparison. These two represent the times of maximum (October 14th) and minimum (July 5th) CO for our set of observations. The shape of each continuum is indicative of the temperature, with the spectra from October 14th noticeably cooler than July 5th, and the strength of the CO feature at 2.3-$\mathrm{\mu m}$ follows the temperature trend. It is worth noting the slope of the $K-$region in the July 5th spectrum is less steep than the October 14th, despite the July 5th observation corresponding to a higher effective temperature. This discrepancy is likely due to the calibration star, as the July 4th spectrum, using a different calibration star, has a steeper slope than the October 14th spectrum. In Section \ref{sec:results}, we discuss other potential artifacts from the telluric correction applied to the July 5th observation. Future observations of CP~Cep at similar phases will use more reliable calibration stars to avoid such issues.

Figure \ref{fig:CO-comp} is a zoomed-in view of the 2.3-$\mathrm{\mu m}$ region for these two nights of observation and also features spectra from an eclipsing binary system, IO Ursae Majoris (IO UMa), for comparison of the CO feature. The primary component in IO UMa is a $\delta$-Scuti variable, and the secondary component is a star of spectral type K and the source of the CO seen in the spectra \citep{soydugan13}. They report the secondary to have a $\mathrm{T_{eff}} = 4260 \pm 150$ K and $\mathrm{log}\ g = 2.71 \pm 0.04$. The CO band heads in the CP~Cep July 5th spectra are difficult to distinguish from the continuum.

\begin{table}
	\centering
	\caption{Table of observations of CP~Cep. $T$ is the total exposure time, $\phi$ is the phase, $X$ is the average airmass, and $\Delta X$ is the difference in average airmass between CP~Cep and the standard. The standards used for telluric correction are given by their HIP catalogue designation.}
	\label{tab:obs}
	\begin{tabular}{c c c c c c}
        \hline
        Obs Date   & $T$ & $\phi$&  $X$   & $\Delta X$ & Standard\\
        UTC        & s   & {   } &        & std-obj & HIP\\
        \hline
        2023-07-04 & 120 & 0.832 & 1.371 & -0.0708 & 106329\\
        2023-07-05 & 450 & 0.893 & 1.137 & -0.0342 & 109911\\
        2023-08-27 & 450 & 0.848 & 1.243 & -0.1112 & 109911\\
        2023-10-13 & 360 & 0.476 & 1.104 & -0.0148 & 109861\\
        2023-10-14 & 360 & 0.529 & 1.167 & -0.0061 & 117602\\
        2023-10-15 & 480 & 0.585 & 1.163 & -0.0030 & 117602\\
        2023-10-16 & 360 & 0.641 & 1.159 & -0.0093 & 117602\\
        2023-10-17 & 480 & 0.697 & 1.156 & -0.0051 & 117602\\
        2023-10-18 & 480 & 0.751 & 1.208 & -0.0014 & 117602\\
        \hline
    \end{tabular}
\end{table}

\begin{figure}
\includegraphics[width=\columnwidth]{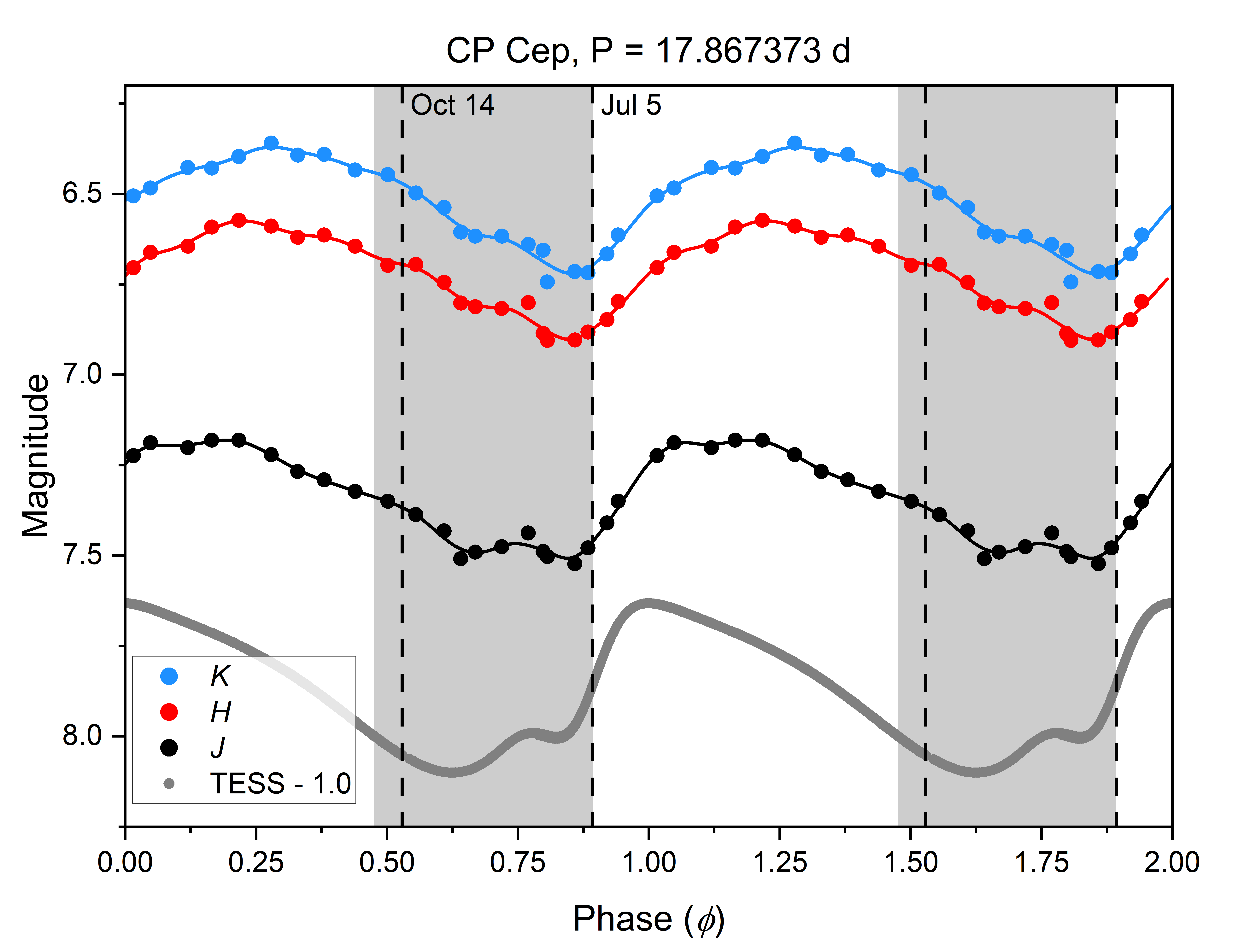}
    \caption{NIR light curves comprised of $JHK$ photometry from \citet{Monson_2011} and the 2022 TESS light curve (shifted 1.5 magnitudes brighter for the purpose of comparison). Our spectral observations fall in the shaded region. The two spectra from Figure \ref{fig:fluxes} are marked by dashed vertical lines.}
    \label{fig:phase}
\end{figure}

Figure \ref{fig:phase} puts our observations in context with the $JHK$ magnitudes from \citet{Monson_2011} with the corresponding \textsc{gloess} fitted light curves \citep{persson04}, as well as the light curve from TESS in 2022 which represents  $R$ to $I$ wavelengths. We adopted the period of 17.867373 days from \emph{Gaia} \citep{gaia16,gaia23}, and we used maximum light from the TESS data as the epoch of $\phi=0$. The observations taken in July and August were near minimum in the NIR and on the ascending portion of the TESS light curve. The October observations cover much of the descending portion of the NIR light curve, or through the minimum in the TESS bandpass. CP~Cep has the ``bump'' at the bottom of the ascending portion of the TESS light curve, the phase location of which is a characteristic of classical Cepheids with periods near 17-days. The bump is also seen prior to minimum light in each of the $JHK$ curves.

\section{Results}
\label{sec:results}

\begin{figure*}
    \includegraphics[width = 0.9\textwidth]{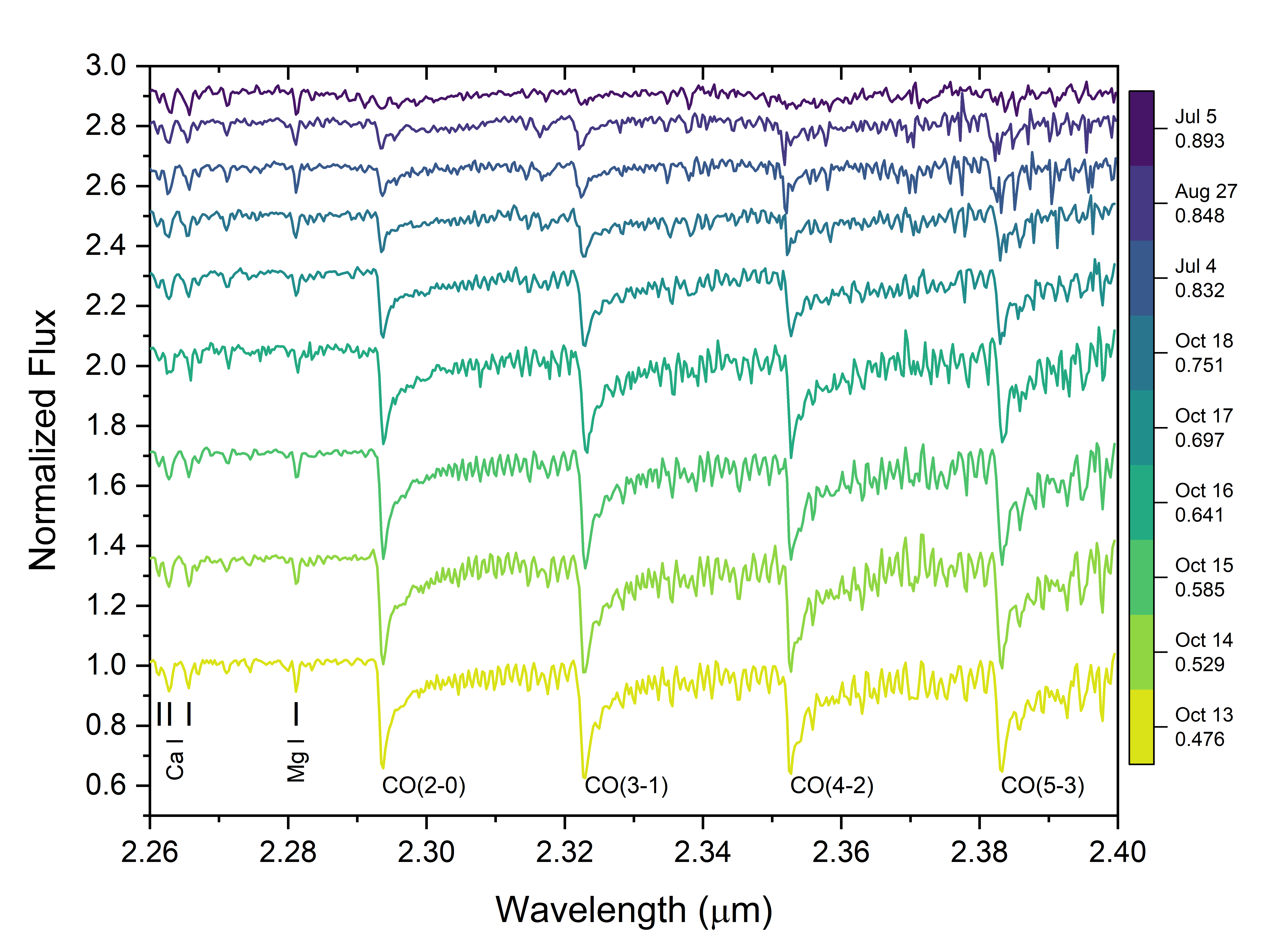}
    \caption{The normalized spectra from each night for a segment of the $K$ region. The observations are organized by increasing phase from bottom to top, and the observations dates and phases are given in the colour bar to the right. Ca and Mg features are labeled as well as the CO band head transitions.}
    \label{fig:KnormCO}
\end{figure*}

\begin{figure}
    \includegraphics[width = \columnwidth]{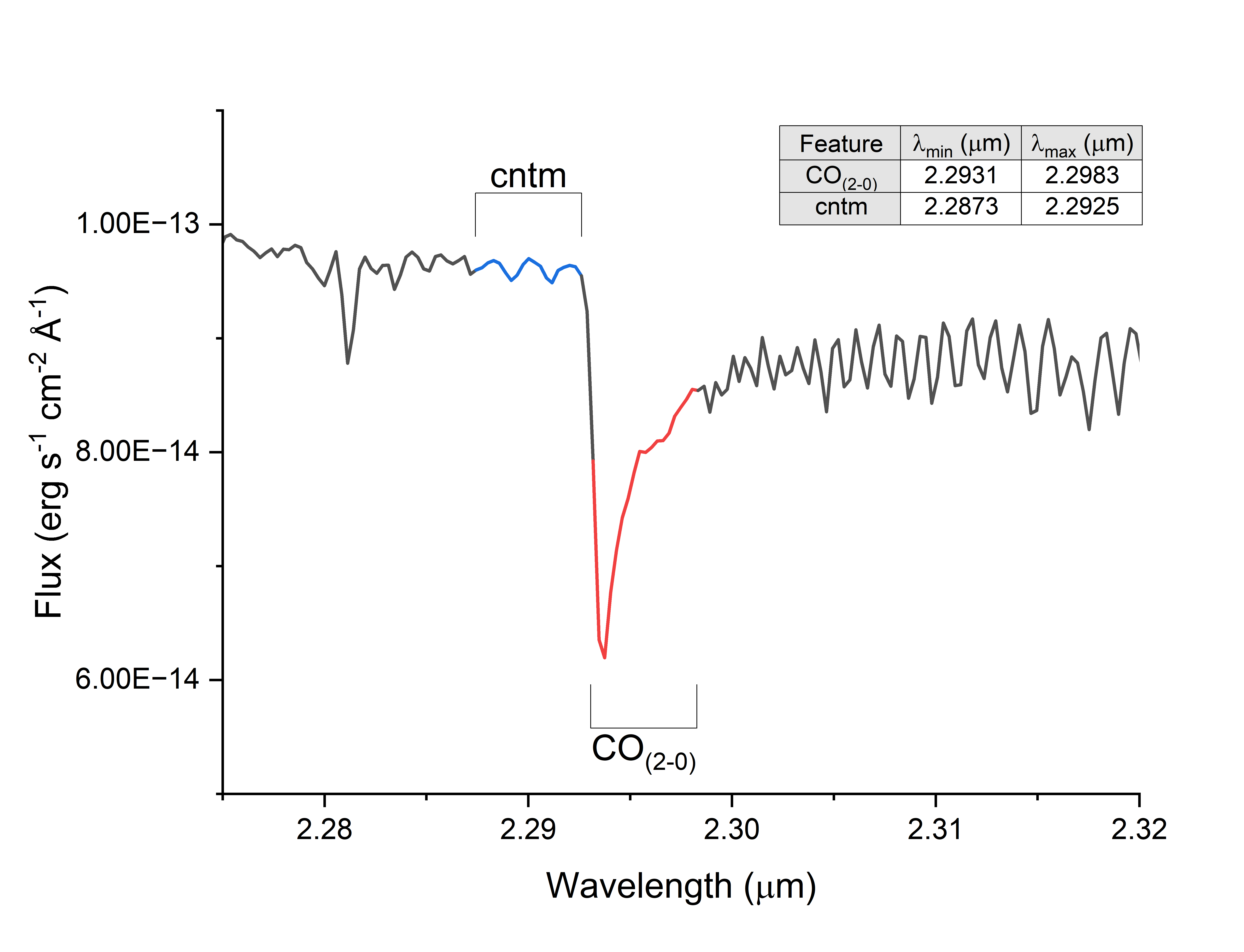}
    \caption{The feature `CO$_{(2-0)}$' region from \citet{kleinmann86} and continuum region from \citet{gonneau16} used for calculation of the CO index overlaid on the October 13th data.}
    \label{fig:COindex_wave}
\end{figure}

The CO absorption for the night of October 14th is strong enough that five band heads are easily distinguishable as in Figure \ref{fig:fluxes}. These correspond to the molecular ro-vibrational energy transitions where $\Delta \mathrm{v} = 2$. Therefore, the band head at the shortest wavelength $\approx 2.29\, \mathrm{\mu m}$ represents the 2-0 transition, the next at $\approx 2.32\, \mathrm{\mu m}$ is the 3-1, and so forth. 

Because CP~Cep is bright in the NIR with a mean magnitude of 7.330 in $J$ \citep{Monson_2011}, the signal-to-noise is high for our dataset. The instrumental pixel uncertainties were typically two orders of magnitude lower than the corresponding fluxes for the individual frames, and combination increased the instrumental SNR further. As discussed in Section \ref{sec:observations}, telluric correction near the order edges is poor and the SNR decreases significantly. In this dataset, the 5-3 and 7-5 transitions are affected by the poor telluric correction, therefore we focus our analyses and discussion on the first three transitions as a result. The instrumental SNR for the 2-0 band head region was above 100 for each observation. Instead of adopting these values, we measured the SNR using the \texttt{splot} task in \textsc{iraf} \citep{tody86,tody93} which computes the SNR by dividing the mean flux by the root mean square. We measured the SNR in the continuum region near to the 2-0 band head ($2.287-2.292 \,\mathrm{\mu m}$), and those measurements are given in Table \ref{tab:index}.

The normalized CO region spectra from each observation are shown in Figure \ref{fig:KnormCO}. These are organized by increasing phase from the bottom to the top. For the earliest phase observations, the band heads are deep and the rotational lines following the band heads are easily distinguished especially those after the 2-0 transition. The later phases show the depths of the band heads and lines decreasing, corresponding to an increase of temperature. By the July 5th observation, the feature is near continuum level. The July 5th data shows the even transitions, 2-0 and 4-2, becoming very broad, appearing as a wide dip in the continuum while the 3-1 transition retains its shape as the depth decreases. The closest observation in phase to July 5th, August 27th, shows hints of this broadening as well. A different standard star was used for calibration on those nights than the other observations. We cannot rule out that the broadening is connected to the telluric correction process using that standard, though we have yet to identify anything in the standard spectra that could be the cause. Two metal features are marked in Figure \ref{fig:KnormCO}; the Ca triplet near 2.26 $\mathrm{\mu m}$ and the Mg I line at 2.281 $\mathrm{\mu m}$. 

Historically, the most common transition to measure CO strength in the NIR is the 2-0 band head. The $^{12}$CO(2,0) index from \citet{kleinmann86} has been frequently used for late-type stars. \citet{gonneau16} adopted this index but measured the continuum as immediately near the first band head. This is the same continuum region that we used for estimates of the SNR discussed previously. For our measurements of CO, we used the same regions as \citet{gonneau16}, and a visualisation of these is shown in Figure \ref{fig:COindex_wave}. We calculated the CO index, $\mathrm{I_{CO}}$, using
\begin{equation}
    \mathrm{I_{CO}} = -2.5 \, \mathrm{log_{10}}\left(\frac{F_{\mathrm{CO_{(2-0)}}}}{F_{\mathrm{cntm}}} \right)
\end{equation}
where $F_{\mathrm{CO_{(2-0)}}}$ and $F_{\mathrm{cntm}}$ are the integrated fluxes for the feature and continuum regions respectively. Associated errors for the index were calculated using the SNR for the continuum region found using \textsc{iraf}. Values for $\mathrm{I_{CO}}$ and errors are shown in Table \ref{tab:index}. 

\begin{figure*}
	\includegraphics[width = 0.9\textwidth]{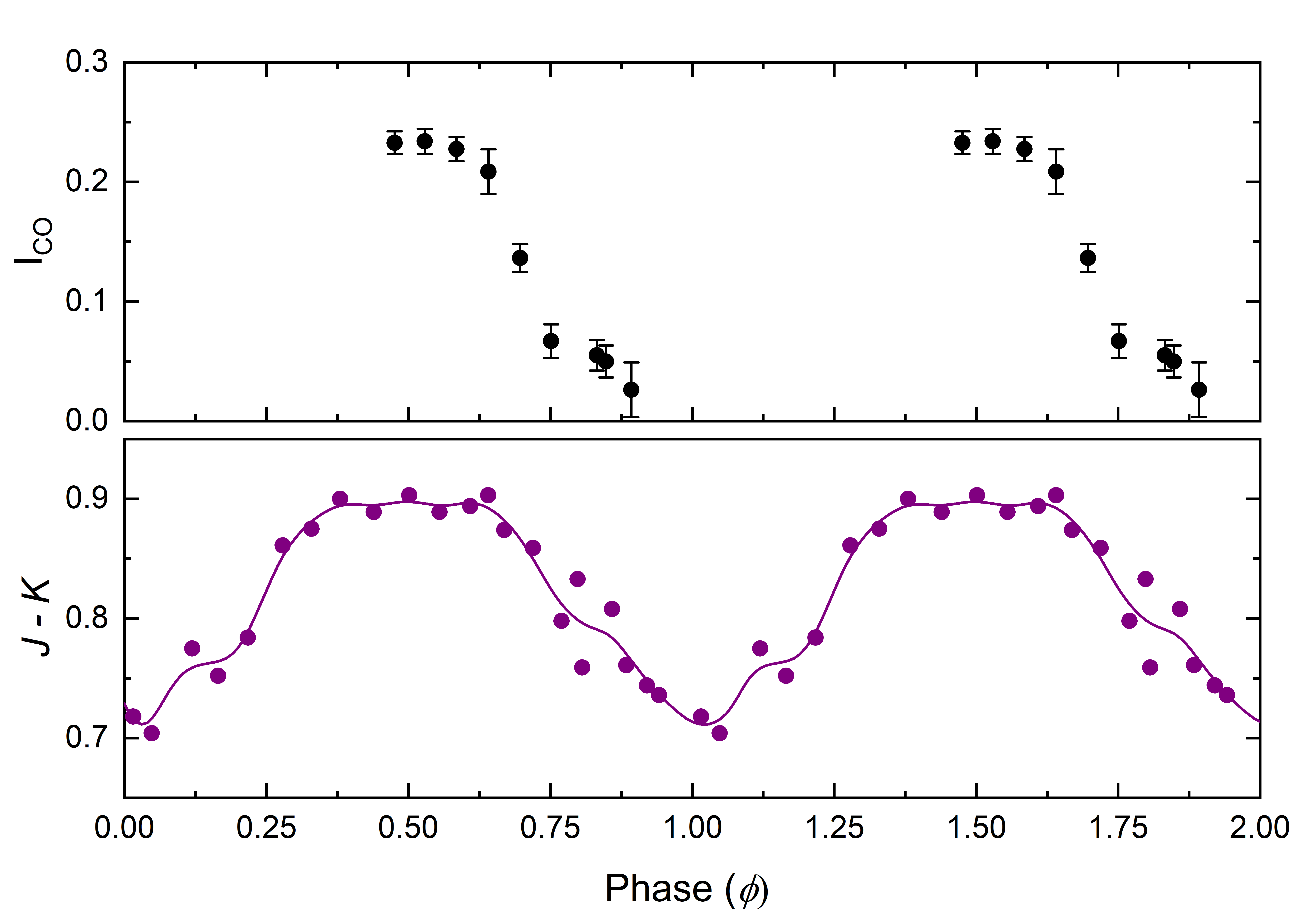}
    \caption{The calculated I$_{\mathrm{CO}}$ values from Table \ref{tab:index} for our observations (top) and the $(J-K)$ colour from \citet{Monson_2011} with corresponding \textsc{gloess} curve (bottom).}
    \label{fig:COindex}
\end{figure*}

Figure \ref{fig:COindex} shows the $\mathrm{I_{CO}}$ values plotted with phase. The $(J-K)$ colour values from \citet{Monson_2011} and corresponding \textsc{gloess} curve is also shown. The $(J-K)$ colour maximum is a plateau for roughly a third of the phase. This represents the star at minimum temperature, and we see this reflected in the values for $\mathrm{I_{CO}}$. The CO values also form a plateau, then sharply decrease for $0.60 < \phi < 0.75$. At $\phi > 0.75$ the dissociation rate of CO decreases. This matches the brief change in slope in the $(J-K)$ curve that is caused by the bump seen in Figure \ref{fig:phase}.

\begin{table}
	\centering
	\caption{Values for the CO index, $\mathrm{I_{CO}}$, and SNR estimates. Table is organized chronologically, not by phase.}
	\label{tab:index}
	\begin{tabular}{c c c}
        \hline
        {Obs Date} & CO$_{(2-0)}$ & SNR \\
        \hline
        2023-07-04 & $0.055 \pm 0.013$ & 120  \\
        2023-07-05 & $0.026 \pm 0.023$ & 67   \\
        2023-08-27 & $0.050 \pm 0.013$ & 115  \\
        2023-10-13 & $0.233 \pm 0.010$ & 161  \\
        2023-10-14 & $0.234 \pm 0.011$ & 146  \\
        2023-10-15 & $0.227 \pm 0.010$ & 152  \\
        2023-10-16 & $0.209 \pm 0.019$ & 82  \\
        2023-10-17 & $0.136 \pm 0.012$ & 132  \\
        2023-10-18 & $0.067 \pm 0.014$ & 110  \\
        \hline
    \end{tabular}
\end{table}

The 2.3-$\mathrm{\mu m}$ CO feature falls within the $K$ and $K_s$ bandpasses in differing amounts. For example, the $K$ bandpass from the Mauna Kea Observatory (MKO) filter set covers the first three transitions (2-0, 3-1, 4-2) at nearly full transmission \citep{tokunaga02}. The MKO $K_s$ filter captures the 2-0 band head at approximately 60\% and the 3-1 band head at effectively 0\% transmission. The $K_s$ filter was developed to minimize the effect of the atmosphere's thermal background for ground-based telescopes, and \citet{persson98} showed that the CO band head was responsible for the scatter seen in $(K - K_s)$ for red stars. 

\begin{figure}
\includegraphics[width = \columnwidth]{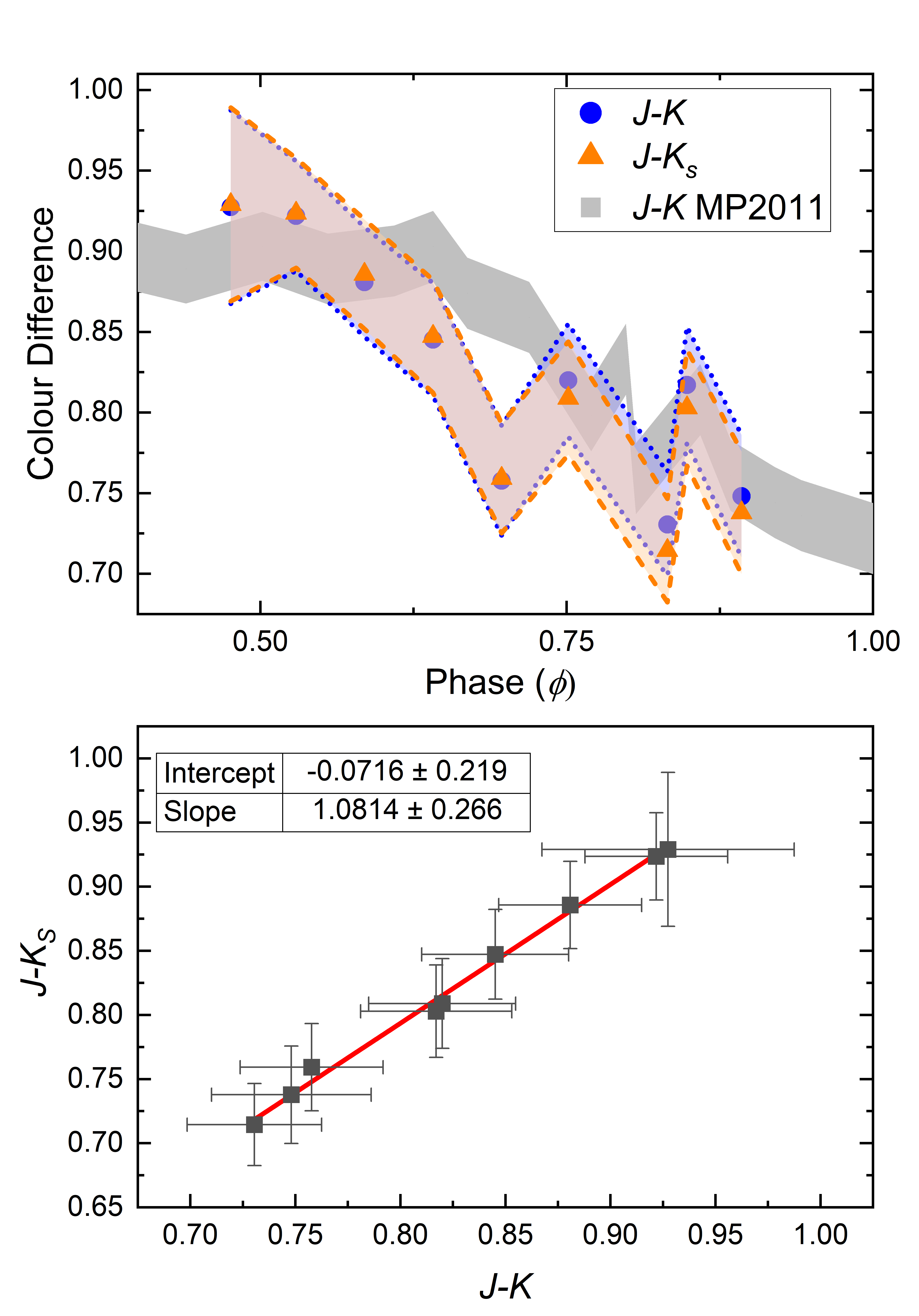}
    \caption{Above: The colour differences, $(J - K)$ and $(J - K_s)$, from the spectra convolved with the Mauna Kea Observatory filter profiles. The shaded regions represent the error bars. The gray shaded region represents the \citet{Monson_2011} errors for comparison. Below: the convolved $(J - K_s)$ versus $(J - K)$ with a weighted linear fit. The slope and intercept of the line are given in the table in the top left.}
    \label{fig:filters}
\end{figure}

To explore the effect of CO on $K$ and $K_s$ magnitudes for CP~Cep, we convolved our spectra with the MKO filter profiles to obtain $J$, $K$, and $K_s$ magnitudes. We used the same process to find the instrumental, convolved magnitudes for the standard stars and found zeropoints using values from 2MASS \citep{cutri2}. We then corrected the CP~Cep $(J-K)$ and $(J-K_s)$ colours using the colour differences from the corresponding standards. The top panel of Figure \ref{fig:filters} shows the results in the form of a colour difference plot, where $(J-K)$ and $(J-K_s)$ are plotted with phase and compared with the values from \citet{monson12}. These convolved colours will not be as accurate as photometric observations and they are intended only to probe the differences between $K$ and $K_s$, but our values match those from the literature relatively well.

When finding the convolved magnitude zeropoints via the standards, we used the 2MASS NIR magnitudes from \citet{cutri2}. Because the standard stars are of spectral type A0, we made the assumption that the 2MASS $K_s$ magnitude is equal to the MKO $K$ and $K_s$. In the case of the standard used for the October 13th observation, the 2MASS $K_s$ magnitude is flagged as an upper limit due to poor photometry. In this case, we assigned the $K_s$ magnitude to be equivalent to the 2MASS $J$ magnitude and implemented a generous error. 

The bottom panel of Figure \ref{fig:filters} shows the convolved $(J-K_s)$ plotted against $(J-K)$. The slope of the line being greater than one indicates a difference between $K$ and $K_s$ based on the phase. This linear fit is weighted for the uncertainties and a slope of one is well within the error range. In the top panel we see that the colours slightly deviate at later phases where the magnitude in $K < K_s$, but again, within calculated errors. This is as the star is approaching maximum temperature and CO is being destroyed, however, the largest difference between $K$ and $K_s$ is only 0.016 dex.

To estimate the amount of light blocked by the CO bands, we replaced the CO band regions ($> 2.29$ $\mathrm{\mu m}$) with continuum values for each observation and performed the MKO $K$ and $K_s$ filter convolutions. For the strongest CO absorption, observed between October 13-17, the $K$-band was, on average, 0.014 magnitudes brighter when the CO region was replaced with continuum values, while the $K_s$-band was 0.003 magnitudes brighter. The region replaced by continuum includes absorption from atomic species as well, so these magnitude differences are larger than if only the CO absorption bands were removed. For context, the $K$ photometric errors for CP~Cep observations from \citet{Monson_2011} were 0.017 mag.

\section{Discussion}
\label{sec:discussion}

\begin{figure}
	\includegraphics[width = \columnwidth]{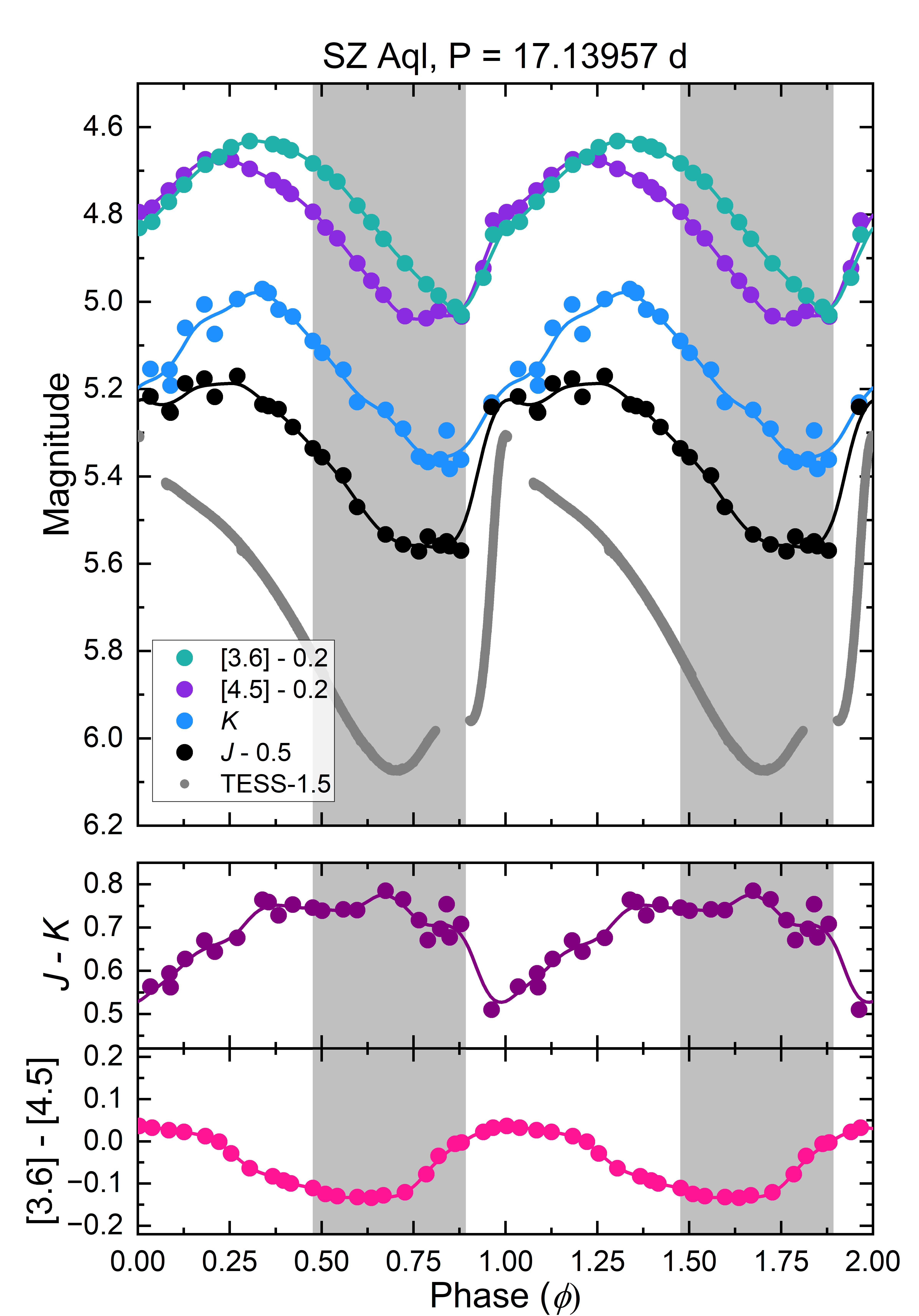}
    \caption{Light and NIR+MIR colour curves for SZ~Aql, a Galactic Cepheid with a similar period to CP~Cep. $J$ and $K$ data from \citet{Monson_2011} and MIR observations from \textit{Spitzer} \citep{monson12,scowcroft16}. $\phi=0$ corresponds to maximum light in the TESS observations. The phases of the CP~Cep observations presented in Section~\ref{sec:observations} are indicated by the shaded region.}
    \label{fig:sz_aql_comparison}
\end{figure}

Figure \ref{fig:sz_aql_comparison} shows the optical (TESS 2022), NIR \citep[$J$ and $K$;][]{Monson_2011} and MIR \citep[\textit{Spitzer}-IRAC;][]{monson12, scowcroft16} light curves (top panel), the $(J-K)$ (middle) and MIR colour curve (bottom panel) for SZ~Aquilae (hereafter SZ~Aql). SZ~Aql is a Galactic Cepheid with similar period to CP~Cep ($\Delta P < 0.05\%$), with the advantage of having well-sampled 3.6 and 4.5~$\mu$m light curves from \textit{Spitzer} available. As the two Cepheids have approximately the same period, they should have very similar mean-light colours (i.e. approximately equal $T_{\text{eff}}$), and should exhibit similar variations in CO. Hence, we can use the MIR light curve of SZ~Aql as a model for CP~Cep. 

The observations of CP~Cep in Figure~\ref{fig:KnormCO} and the measurements in Figure~\ref{fig:COindex} show clear evidence of varying CO, with the strongest CO features occurring at the largest $(J-K)$ colour difference or coolest temperature. This is consistent with the behaviour seen in the MIR colour curve of the SZ~Aql model in Figure~\ref{fig:sz_aql_comparison}. In the MIR colour difference plot, the effect of CO absorption is the strongest near $\phi \approx 0.6$, beyond which dissociation begins and the colour difference approaches 0. 

Our direct spectroscopic measurements of CO in CP~Cep parallel the photometric effect seen in the MIR for SZ~Aql. Assuming similar behaviour in CP~Cep and SZ~Aql, our earliest phase observation would occur just before the minimum of the MIR colour curve; close to maximum CO absorption in the atmosphere. After this, the CO dissociation rate starts off gradual and then rapidly increases as the Cepheid approaches maximum $T_{\text{eff}}$. Our latest phase observation would fall just before the MIR colour curve begins to flatten off, where CO has mostly been destroyed due to high temperature.

The stellar atmospheric modelling presented in \cite{scowcroft16} used the ATLAS9 1-D Local Thermodynamic Equilibrium (LTE) model to predict CO variation across the temperature range expected of Cepheids. The LTE approximation is validated by simulation work by \cite{Ayres&Wiedemann1989} who concluded that the MIR spectrum of CO in cooler stars (such as Cepheids) is indeed close to the LTE ideal. Advances in stellar atmospheric modelling using non-LTE models have recently resulted in differences in synthetic spectra for some chemical species \citep{Bergemann2017}. For simple molecules and radicals, this difference in behaviour between LTE and non-LTE models is driven by radiative over-dissociation \citep{Popa2023}, and is therefore expected be important for species with low dissociation energies eg: OH, CH, NH, H\textsubscript{2}O, but not for those with high dissociation energies such as CO. Where non-LTE modelling may be relevant to the observed behaviour of Cepheids in the MIR is in relation to the turnover in the plot of MIR colour versus log P (see \cite{scowcroft16}). This unexpected behaviour suggests there is an enhanced rate of destruction of CO at the coolest temperatures experienced in Cepheid atmospheres.  The over-dissociation, under non-LTE conditions, of species participating in the CO reaction schemes (or their precursors) could account for this by shifting the position of equilibrium in one or more of the reactions, but more work is needed to understand this complex behaviour. The role of 3-D effects on CO features in the spectra of late-type stars may also be relevant to the understanding of this behaviour \citep{Dobrovolskas}.

The agreement between the NIR CO spectra and the predictions from the MIR colour curve of a similar Cepheid provides strong evidence that the mechanism for cyclical CO formation and destruction proposed in \citet{scowcroft16} holds true, and hence that CO is behind the variation in mean MIR colour for Cepheids. 

Scowcroft et al.'s idea was that using mean MIR CO colour as a proxy for CO, and using CO as a proxy for metallicity, would allow a low cost means of obtaining metallicities for Cepheids and hence the PCZ relation for these stars.  \citet{takeda13} showed a moderate underabundance of C ([C/H] -0.3 dex), in a small sample of low-period, solar metallicity Cepheids, this being interpreted as the canonical dredge-up of CN-cycled material, while no such effect was demonstrated for [O/H]. These findings have since been supported by \citet{luck18}, and suggest the relationship between the observed metallicity of Cepheids and the atmospheric abundances of C and O is not straightforward.

There remains a debate over the role of metallicity in the Cepheid distance scale, and its potential contribution to the current Hubble tension \citep{freedman21}. If the ability to use MIR colour as a proxy for metallicity is confirmed, then \textit{JWST}-NIRCam photometry of Cepheids in Local Group galaxies, combined with spectroscopic metallicities, will enable calibration of the MIR PCZ. This will give access to accurate metallicities for individual Cepheids in type Ia supernova host galaxies, rather than relying on indirect metallicity attribution from H II regions, and help to settle this debate.

\section{Conclusions}
\label{sec:conclusions}

We obtained NIR spectra for the classical Cepheid, CP~Cep, over more than 40\% of the pulsation cycle. The 2.3-$\mu$m CO feature is present in each spectra. The strengths of the CO band heads change with the pulsation of the star, and match well with the $(J-K)$ colour curve. Comparison of NIR spectroscopic observations of CP~Cep to the MIR photometric observations of a similar Galactic Cepheid, SZ~Aql, provides further evidence of the CO dissociation process explored in previous work \citep{scowcroft16}. Unlike the [4.5]-$\mu$m \textit{Spitzer}-IRAC bandpass, the CO absorption in the NIR does not significantly affect magnitudes in the $K$ region, and therefore the period-luminosity relation in $K$ is not impacted by the NIR CO bands. 

The spectral observations of CP~Cep occurred during three different pulsation cycles, and from initial to most recent observation was almost six pulsation cycles. The continuity of measurements of CO destruction throughout these further confirm that the CO absorption is in the star and not the interstellar medium.

\section*{Acknowledgements}
We thank the anonymous reviewer for their insightful comments and suggestions, which have improved the quality of this work. This paper includes data collected by the TESS mission, which are publicly available from the Mikulski Archive for Space Telescopes (MAST), and observations made with the Spitzer Space Telescope, which was operated by the Jet Propulsion Laboratory, California Institute of Technology under a contract with NASA. Based on observations obtained with the Apache Point Observatory 3.5-meter telescope, which is owned and operated by the Astrophysical Research Consortium. VS thanks Dr David Carty for fruitful discussions on CO chemistry.

\section*{Data Availability}

The data discussed in this paper will be made available upon reasonable request.



\bibliographystyle{mnras}
\bibliography{example} 





\bsp	
\label{lastpage}
\end{document}